\documentclass[aps, prl, twocolumn, reprint, amsmath]{revtex4-1}

\usepackage{color}
\usepackage{graphicx}
\usepackage{bm}
\usepackage{amsmath, mathrsfs}
\usepackage{amsbsy}

\begin{document}

\title{Asymmetry in laser wakefields driven by intense pulses}

\author{Zs. L\'ecz}
\email{Zsolt.Lecz@eli-alps.hu}
\affiliation{ELI-ALPS, ELI-HU Nonprofit Ltd. Wolfgang Sandner u. 3, H6728 Szeged, Hungary}

\author{Sz. Majorosi}

\affiliation{ELI-ALPS, ELI-HU Nonprofit Ltd. Wolfgang Sandner u. 3, H6728 Szeged, Hungary}

\date{\today}

\begin{abstract} 
Laser wakefield theories rely on the laser envelope function, which is radially symmetric, and predict zero transverse momentum for the electrons along the propagation axis. Exact description of laser wakefields, beyond the envelope approximation, requires a more general formula for the Lorentz force acting on the electrons. Here we present a fundamental approach to express the transverse momentum of an electron crossing the laser pulse and we show that an exact analytical formula can be derived for the non-zero transverse momentum of electrons initially laying along the axis of symmetry. The results outlined here shed light on the details of the electron motion inside an intense laser pulse and explain the strong wakefield asymmetry observed in simulations.
\end{abstract}

\maketitle

Laser-driven electron acceleration in low density (gaseous) plasma is based in the fact that electrons are displaced longitudinally, in the propagation direction of the laser, while the ions remain practically immobile \cite{Gibbon}. This charge displacement gives rise to a longitudinal electric field and consequent plasma oscillation, which forms a plasma wave with phase velocity equal to the linear group velocity of the laser pulse. Upon injection of electrons into this alternating field structure, some of them can reach a velocity close to the speed of light and get accelerated to relativistic energy. Since the acceleration takes place behind the driving pulse, in its wake, this scheme is named as Laser Wakefield Acceleration (LWFA) \cite{tajima1979}, and has been extensively investigated in the last three decades. In all fundamental theories describing the injection or acceleration stages a symmetric transverse laser envelope is considered \cite{Esarey, selfFocus}, therefore the wake structure is also symmetric in the radial direction. Furthermore, the electron motion in the laser field is also simplified, by removing the oscillating part connected to the laser frequency, and an averaged trajectory is considered which is governed by the laser envelope only.

Due to the radial symmetry in the laser intensity and in the electron motion, the whole interaction can be usually modeled in the envelope, or ponderomotive, approximation, which is valid only for relatively long pulses. As the laser field amplitude increases or the pulse length decreases the sub-cycle electron dynamics cannot be neglected. Advanced numerical plasma simulation codes can handle all the nonlinearities related to such motion, but these modeling tools are computationally expensive and cannot provide a full understanding of the process. 

The asymmetric scattering of charged particles in a laser field is relevant for indirect laser diagnostics as well, where the detection of the electron's angular distribution can be used to estimate the laser field amplitude in focus \cite{Olga}. This happens in rarefied gas (high vacuum) where the plasma fields can be fully neglected and the particle motion is perfectly symmetric in the case of ions, because their transverse displacement is very small \cite{OlgaProton}. 

Although the generation of few-cycle laser pulses above 1 J energy level is still challenging, in a high-density plasma \cite{expSPM} single-cycle waveforms can appear even in a long pulse due to strong self-phase modulation, which is a highly non-linear effect \cite{asymCompress, pulseSteep}. This phenomenon is responsible for bubble wobbling \cite{bubbleWobbling} and for the consequent electron beam pointing fluctuations \cite{BeamPoint}. In the case of a short acceleration distance the electron beam pointing can be well controlled by setting the initial carrier envelope phase (CEP) of the pulse \cite{Wcontrol}, but in the case of a longer distance the CEP changes during the propagation and in the nonlinear regime it has a strong impact on the acceleration \cite{FewCycleSelfInj, identifyCEP}. Therefore, understanding the  wakefield formation and the electron motion in ultrashort pulses is crucial in many experimental scenarios, where the control over the electron beam parameters is highly challenging.

In this letter we present a semi-analytical framework suitable to predict the asymmetry in the electron distribution in the case of laser pulses with relatively large focal spot diameter. A similar attempt was presented in Ref. \cite{Carrier} using perturbation theory, but, as we show here, an exact equation for the tranvserse momentum can be derived usig a more simple approach. Furthermore, we show that the longitudinal component of the laser electric field does not have a strong contribution to the asymmetry, which is the opposite of what was assumed in Ref. \cite{Carrier}.

The position $\mathbf{r}(t)$ and momenta $\mathbf{p}(t)$ of any electron in a laser pulse can be calculated from the the equations of motion with the Lorentz-force expressed with $\phi(x,y,z,t)$ scalar and $\mathbf{a}(x,y,z,t)$ vector potentials as \cite{Landau}:
\begin{equation}\label{eq:motion} 
\frac{d(\mathbf{p-\mathbf{a}})}{dt} = \nabla \phi -\mathbf{v} \times \nabla \times \mathbf{a} - (\mathbf{v}\cdot \nabla)\mathbf{a},
\end{equation}
where $\mathbf{a} = \mathbf{a}(\mathbf{r}(t),t)$ and the derivatives of $\phi$ and $\mathbf{a}$ are evaluated at the particle position $\mathbf{r}(t)$. The laser is represented in the transverse component of the vector potential as $\mathbf{a} = a_x \mathbf{k}+a_y \mathbf{j}$ and propagates in the $z$ direction. We normalize time and spatial coordinates to one over the laser's carrier angular frequency ($\omega_0$) and wave vector; and $a$, momenta, velocities to $m_ec/e$, $m_ec$ and $c$, respectively.
The velocity is expressed as $\mathbf{v}=\mathbf{p}/\gamma$, where $\gamma=1/\sqrt{1-|\mathbf{v}|^2}=\sqrt{1+|\mathbf{p}|^2}$ is the relativistic Lorentz factor associated to the motion. The evolution of the momentum components in a laser field is highly nonlinear, because the laser's potential is evaluated at the position of the electrons, which in turn depends in the momentum $d\mathbf{r}/dt=\mathbf{p}/\gamma$. The longitudinal component of the laser's electric field can be represented by a scalar potential ($\phi$), which can include the potential of the wakefield as well, if plasma fields are also considered.

From the principle of work done on an object the energy rate equation can be derived:

\begin{equation}\label{eq:energyrate}
\frac{d\gamma}{dt} =\mathbf{v}\cdot \left(\nabla \phi + \frac{\partial \mathbf{a}}{\partial t}\right).
\end{equation}
In the followings we consider an analytical form of the vector potential written as $a(x,z,t)=\tilde a \cos(t-z/v_p+\varphi_0)$, where $\tilde a=a_0\sin[(t-z)/(2N)]^2\exp(-x^2/w_0^2)$, $N$ is the number of laser cycles in the laser pulse, $w_0$ is the waist radius and $\varphi_0$ is the CEP offset. The phase velocity, in the paraxial approximation \cite{nonParax}, is given as $v_p=(1+2/w_0^2)/(1-2/w_0^2)$, which is valid close to the laser's axis of symmetry (where $x\ll w_0$). Combining Eqs. (\ref{eq:motion}, \ref{eq:energyrate}), and using $v_p=1$, yields an important relation, called as the constant of motion:

\begin{equation}\label{eq:constm}
\gamma - 1=p_z+\Phi,
\end{equation}
where $p_z$ is the longitudinal momentum component and $\Phi=\int (1-v_z)(\partial \phi/\partial z)dt$. We note here that the longitudinal field component can be represented by the vector potential as well in the Coulomb gauge, and in that case $\Phi=a_z$. Independently from the laser field amplitude, polarization or focusing geometry, Eq. (\ref{eq:constm}) always holds. Assuming a linearly polarized laser pulse, with a single $a=a_x$ component, the equation of motion in the transverse direction reduces to:

\begin{equation}\label{eq:motionx}
\frac{dp_x}{dt} =\frac{\partial a}{\partial t}+v_z\frac{\partial a}{\partial z},
\end{equation}
which is just the transverse Lorentz force acting on the electrons. On the right hand side one can recognize the total time derivative of the vector potential, therefore a second constant of motion is obtained: $p_x=p_{x0}+a$. However, Eq. (\ref{eq:motionx}) is valid only in planewaves. If the vector potential varies in the transverse direction as well (focused beams) the total derivative is $da/dt=\partial a/\partial t+v_z(\partial a/\partial z)+ v_x(\partial a/\partial x)$. In the case of arbitrary focused pulses we write the transverse component of the momentum as a sum of the planewave solution and a small deviation: $p_x=a+\mathcal{P}$. Substituting this expression of the transverse momentum into Eq. (\ref{eq:motionx}) leads to:

\begin{equation}\label{eq:dpx}
\frac{d\mathcal{P}}{dt} =-v_x\frac{\partial a}{\partial x},
\end{equation}
which is independent of the longitudinal velocity component. If the laser pulse is not tightly focused, then the laser's potential can be expanded around any transverse position as $a(x)\approx a(x_0)+\Delta x \partial a/\partial x+(1/2)\Delta x^2\partial^2 a/\partial x^2+(1/6)\Delta x^3\partial^3 a/\partial x^3=a(x_0)+\Delta xD_1+\Delta x^2D_2/2+\Delta x^3D_3/6 $, where $\Delta x=x-x_0$, $x_0$ is the initial transverse position of an electron at rest before interaction with the laser field and the derivatives are evaluated at this fixed position. In this way the reduced equation of motion in the transverse direction becomes:

\begin{equation}\label{eq:dpx2}
\frac{d\mathcal{P}}{dt} =-\frac{a+\mathcal{P}}{\gamma}[D_1+\Delta xD_2+\Delta x^2D_3/2],
\end{equation}
which is interpreted as the relativistic ponderomotive force. In the case of larger focal spot radii ($D_2, D_3$ are very small) Eq. (\ref{eq:dpx2}) is equivalent to transverse component of $F_p=\nabla \phi-\nabla \gamma$, which is the generalized ponderomotive force with $\gamma=\sqrt{1+p_z^2+(\mathcal{P}+a)^2}$ and in this case $(\nabla \phi)_{\perp}=0$. For clarity, we note here that in the envelope approximation, where asymmetry is fully excluded \cite{MoraP}, the fast component of the electron's momentum (quiver motion) is neglected and the slow component ($\mathbf{p_{s}}$) is used to define an averaged Lorentz factor: $\gamma_s=\sqrt{1+p_{zs}^2+p_{xs}^2+{\tilde a}^2/2}$. In principle, Eq. (\ref{eq:dpx2}) provides a more exact expression for the electron momentum and one can introduce higher-order expansion terms as well, but later we show that three (or even two) terms are enough to precisely reproduce the numerical results of Eq. (\ref{eq:motion}).

\begin{figure}[h]
\centering
\includegraphics[width=0.23\textwidth]{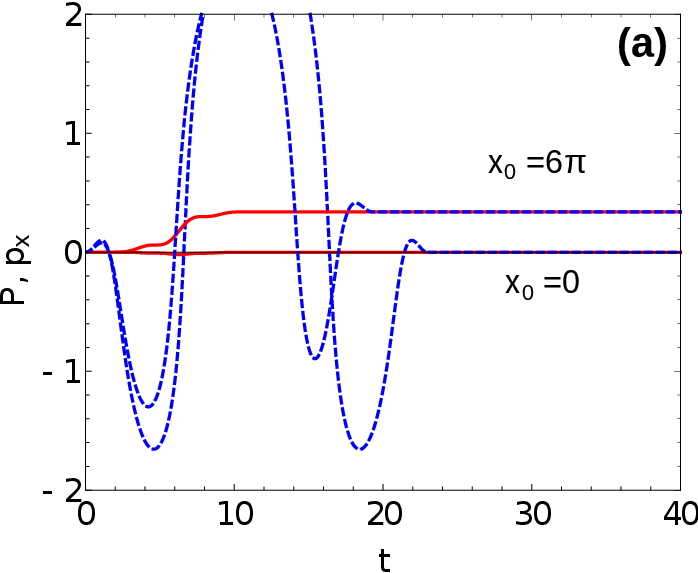}
\includegraphics[width=0.23\textwidth]{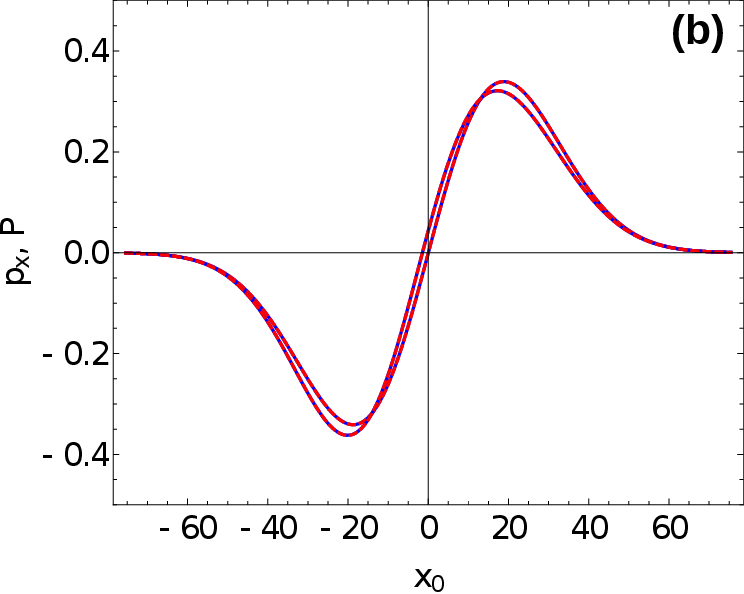}

\caption{ (a) Comparison of $p_x$ (blue, solution of Eq. (\ref{eq:motion})) and $\mathcal{P}$ (red, solution of Eq. (\ref{eq:dpxcoord})) for two values of initial electron position $x_0$. The parameters of the laser pulse are: $w_0=12\pi$, $N=2$ and $a_0=3$. (b) Final values of $\mathcal{P}$ and $p_x$ as a function of initial electron position for two CEP values: $\varphi_0=0$ and $\varphi_0=\pi/2$.}
\label{Fig1}
\end{figure}

In Eq. (\ref{eq:dpx2}) the connection to the longitudinal dynamics appears in the $\gamma$ factor alone, which can be eliminated by introducing the following coordinate transformation: $\tau=t-z$. In this time frame (where $v_z^{\prime}=0$) the time derivative is $d/dt=(d\tau/dt)d/d\tau=(1-v_z)d/d\tau=-(1+\phi)(d/d\tau)/\gamma$, where we used the first constant of motion, Eq. (\ref{eq:constm}), and in this frame of reference $\Phi'=\phi$. We note here that the same coordinate transformation was used in Ref. \cite{Carrier} as well. Furthermore, this choice of coordinate is motivated by the fact that the vector potential in the momentum equation is a function of $\tau$ only, if we consider that the phase velocity of the laser wave is $c$. Thus, in the new coordinate system the equation of motion reduces to a single scalar equation:

\begin{equation}\label{eq:dpxcoord}
(1+\phi)\frac{d\mathcal{P}}{d\tau} =(a+\mathcal{P})[D_1+\Delta xD_2+\Delta x^2D_3/2].
\end{equation}

The electron's tranverse coordinate ($x$) as a function of the proper time ($\tau$) is calculated from the velocity equation $\gamma v_x=(1+\phi)dx/d\tau=a+\mathcal{P}$. In the case of numerical solution the scalar potential is calculated from $\nabla \cdot \mathbf{E}=0$, which can be written as $\partial \phi/\partial \tau=\partial a/\partial x$. Several examples of the numerical solutions are presented in Fig. \ref{Fig1}. It is shown that Eq. (\ref{eq:dpxcoord}) reproduces exactly the final transverse momentum of an electron after crossing the laser pulse. In the case of larger $w_0$ values it is enough to keep only up to the second order term in Eq. (\ref{eq:dpx2}). It is also clearly seen that the transverse momentum is asymmetric with respect to $x=0$ (see Fig. \ref{Fig1}(b)).

\begin{figure}[h]
\centering
\includegraphics[width=0.21\textwidth]{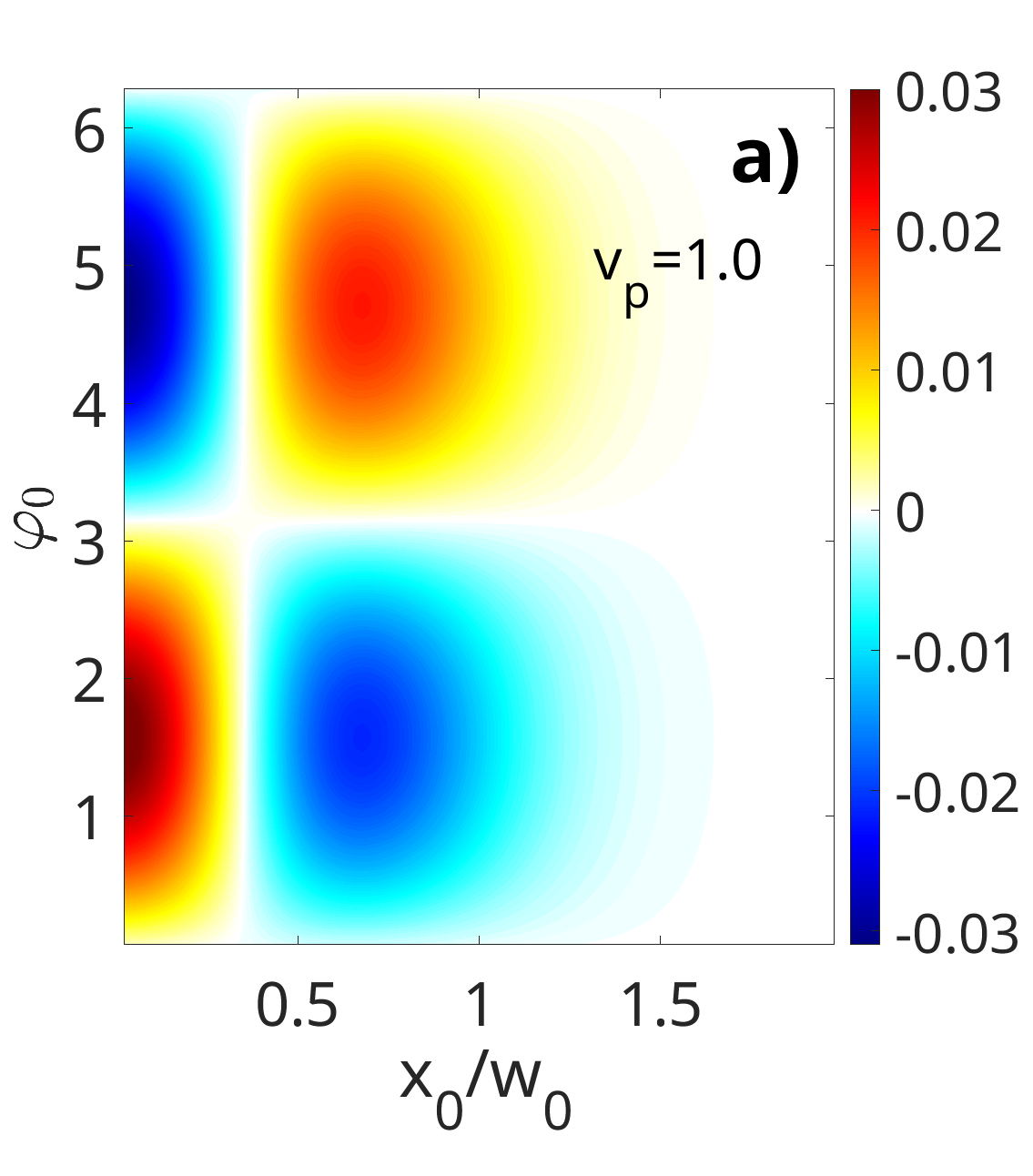}
\includegraphics[width=0.21\textwidth]{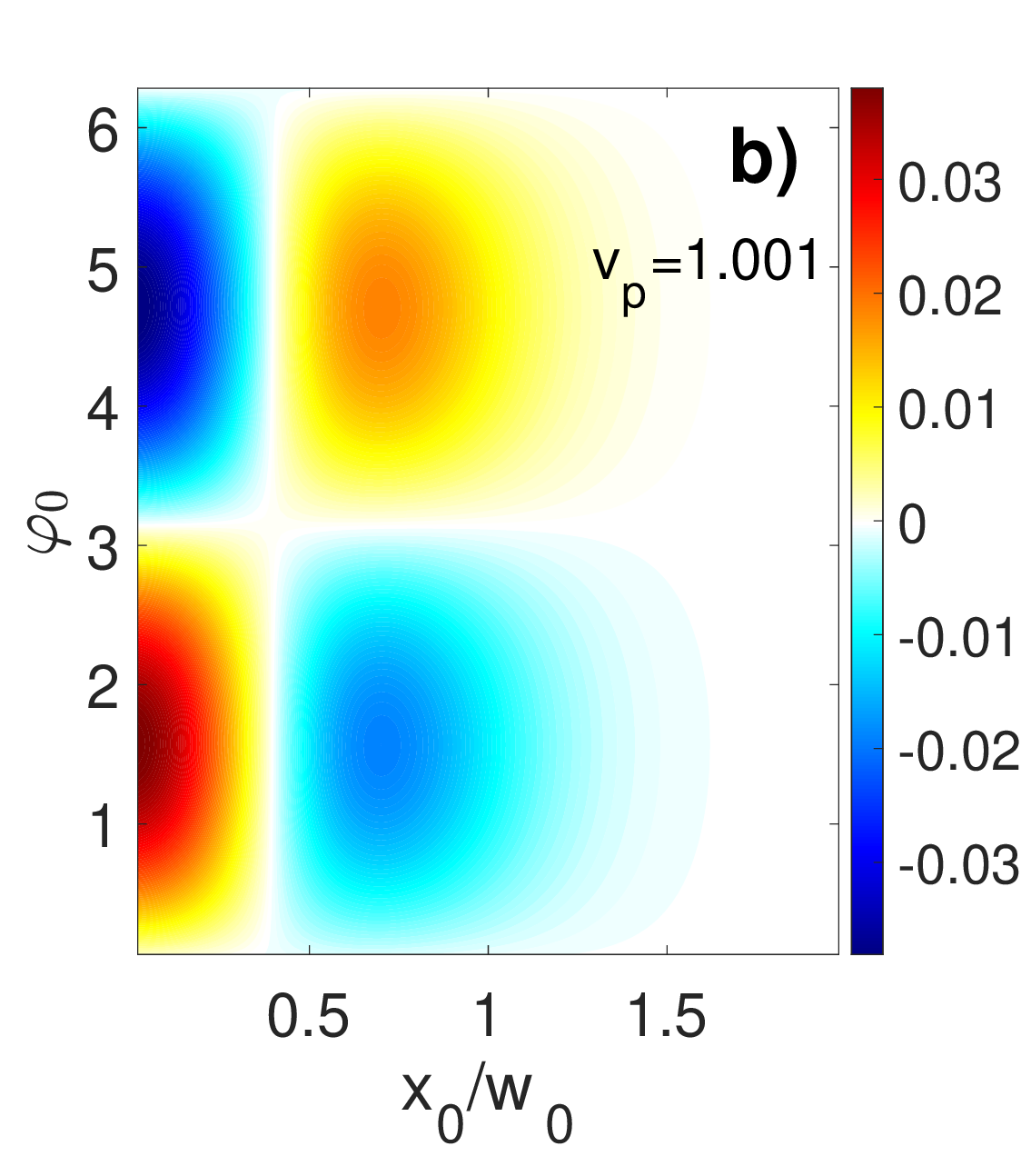}

\includegraphics[width=0.23\textwidth]{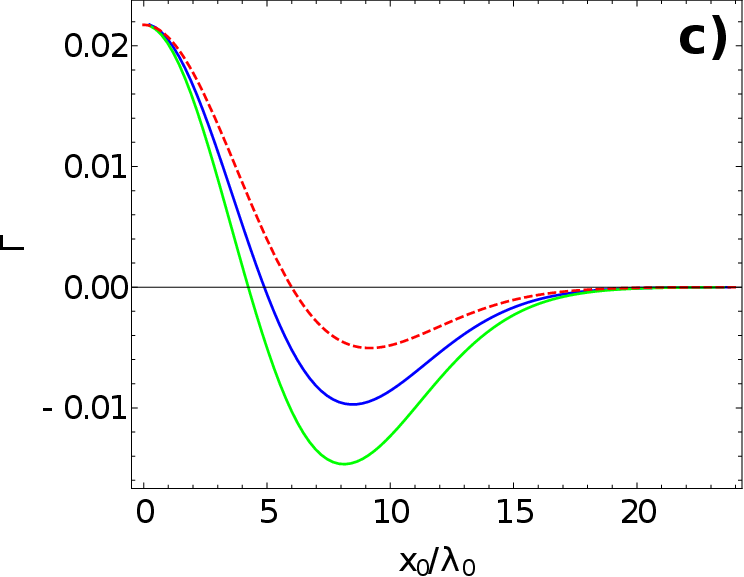}
\includegraphics[width=0.23\textwidth]{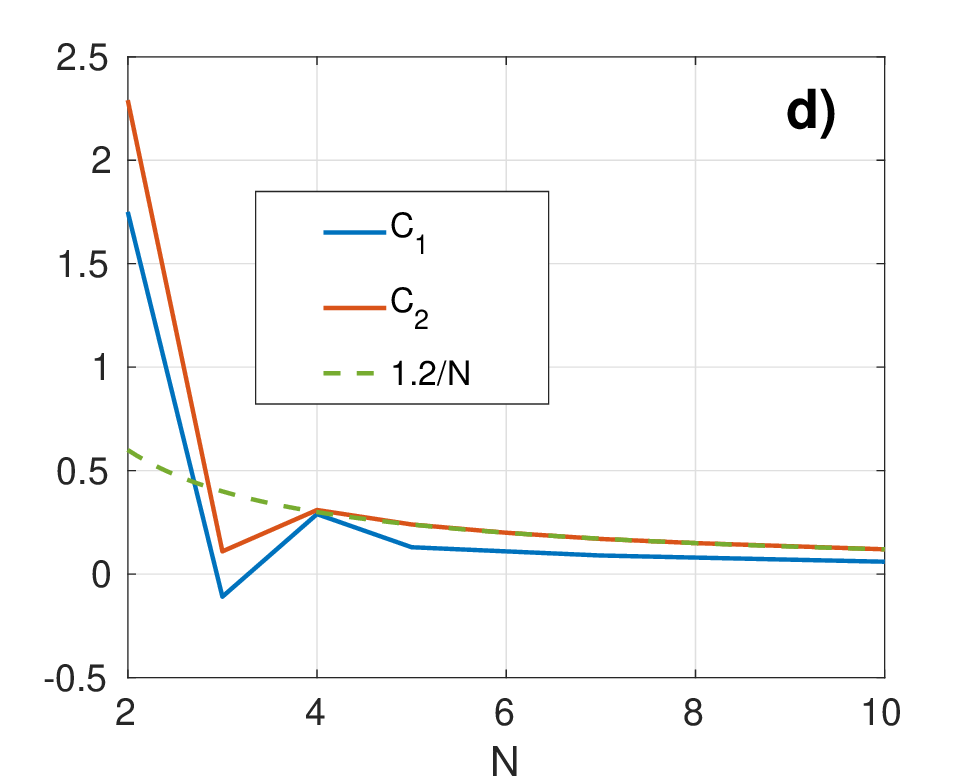}

\caption{ (a) Assymetry parameter ($\Gamma$) for $w_0=24\pi$, $N=2$ and $a_0=3$ calculated with Eq. (\ref{eq:motion}). (b) The same is shown, but with a larger phase velocity. (c) The green line correspond to Eq. (\ref{eq:motion}), with the same parameters as in (a), where $\varphi_0=\pi/2$. The blue line represents the solution of Eq. (\ref{eq:dpxcoord}), with $\phi=0$, and the red dashed line correspond to the analytical solution of Eq. (\ref{eq:dpxcoordsimpl}). In (d) the pulse length dependence of the coefficients used in Eq. (\ref{eq:dpanalit}) are shown. }
\label{Fig2}
\end{figure}

To quantify the asymmetry of a wakefield generated by few-cycle pulses we introduce the variable $\Gamma=\mathcal{P}(-x_0)+\mathcal{P}(x_0)$, that is zero for a symmetric ponderomotive force. We present the exact values of $\Gamma$ in Fig.\ref{Fig2}(a) as a function of transverse coordinate and CEP phase, which shows that the assymetry is the strongest along the axis. Furthermore, this parameter changes sign along the $x$ coordinate for a given $\varphi_0$ value, which is more visible in Fig. \ref{Fig2}(c). We have also tested the effect of superluminal phase velocity, which is shown in Fig. \ref{Fig2}(b). The asymmetry is significantly larger when $v_p$ is increased by 0.1 \% and the radial profile is also slightly different. In the plasma the phase velocity is even larger, therefore the asymmetry is is expected to be larger compared to the prediction of our single-particle modeling.

Fully analytical solution is obtained when $\mathcal{P}$ and $\phi$ are neglected in the right and left hand side of Eq. (\ref{eq:dpxcoord}), respectively, which corresponds to the case of a relatively large focal spot diameter ($w_0\gg \lambda_0$), or near the propagation axis. In this case the momentum equation reduces to (keeping derivatives up to the second order):

\begin{equation}\label{eq:dpxcoordsimpl}
\frac{d\mathcal{P}}{d\tau} \approx [a(x_0)+\Delta xD_1+\Delta x^2D_2/2][D_1+\Delta xD_2].
\end{equation}

Furthermore, we have to assume that the transverse displacement is well approximated with the plane-wave solution, e.g. $\mathcal{P}\ll a$, and $\phi\approx 0$, which results in $dx/d\tau=a(x_0, \tau)$. For a $Sin^2$ temporal intensity envelope the $x$ coordinate is calculated analytically, which is then used in Eq. (\ref{eq:dpxcoordsimpl}). In this way nalytical expression for $\Gamma$ can be obtained, but it is valid only near the axis of propagation, as it is seen in Fig. \ref{Fig2}(c), and for relatively larger laser spots $w_0>a_0$. The transverse momentum of an electron initially located near the axis, after crossing the laser pulse, is:

\begin{equation}\label{eq:dpanalit}
\mathcal{P}(x_0\approx 0) =\mathcal{P}_0= \frac{a_0^3}{w_0^2} \left( C_2 -C_1 \frac{a_0^2}{w_0^2}  \right)\sin \varphi_0,
\end{equation}
where the pulse length dependence of the numerical coefficients $C_1, C_2$ is shown in Fig. \ref{Fig2}(d). We can identify an inverse proportionality, therefore in the case of long pulses the momentum of axial electrons can be expressed as:

\begin{equation}\label{eq:dpanalit2}
\mathcal{P}_{0, N\gg 1} \approx \frac{a_0^3}{w_0^2N}\left(1.2 - 0.6\frac{a_0^2}{w_0^2}\right)\sin\varphi_0,
\end{equation}

The analytical solution in Eq. (\ref{eq:dpanalit2}) is essentially the same as in Ref. \cite{Carrier} (except the $1/N$ dependence), but this result is not restricted to few-cycle pulses and we found a weaker dependence on the puls elength. Asymmetric wakefields are generated by long pulses as well, if $a_0/w_0$ is large enough. However, this is out of the range of validity of Eq. (\ref{eq:dpanalit}), therefore in the tightly focused case we have to fully rely on numerical solutions.

\begin{figure}[ht]
\centering
\includegraphics[width=0.22\textwidth]{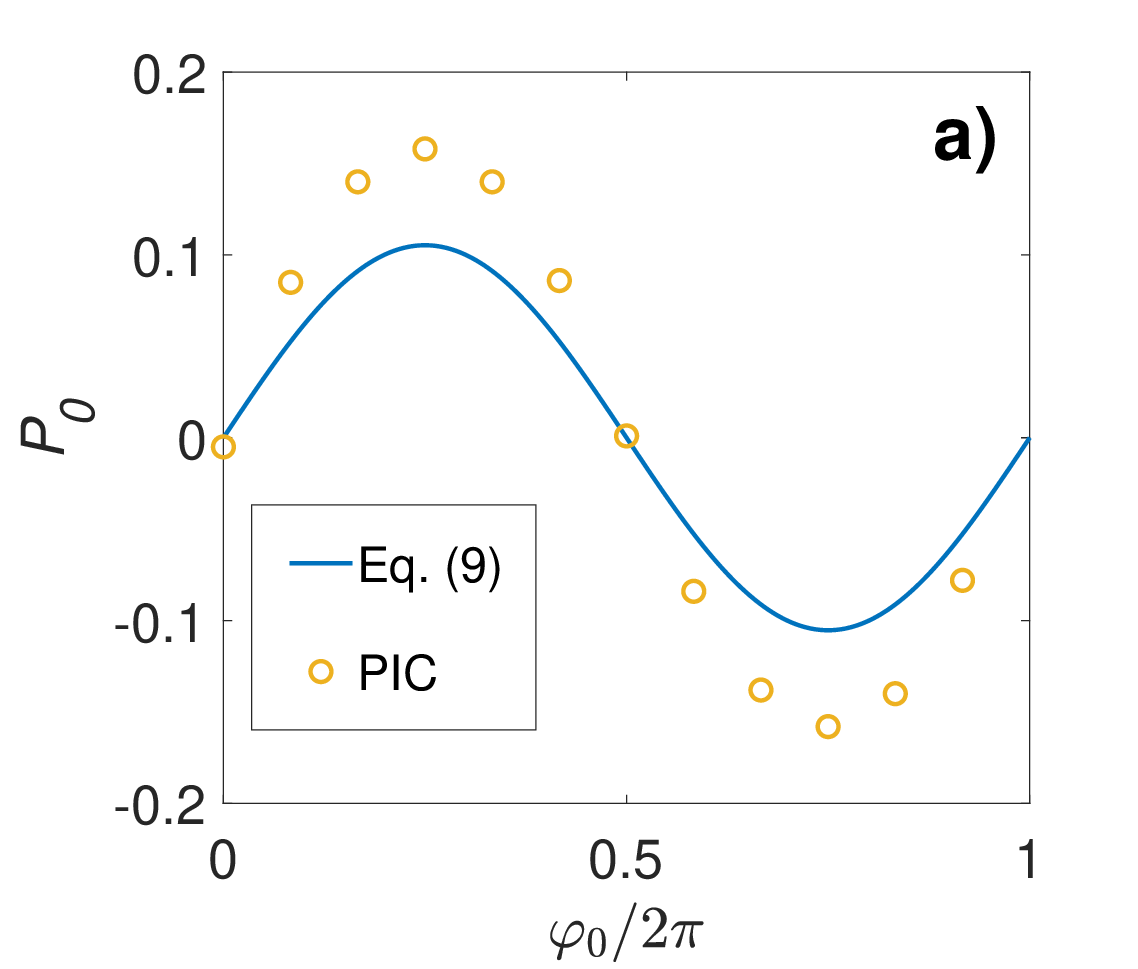}
\includegraphics[width=0.23\textwidth]{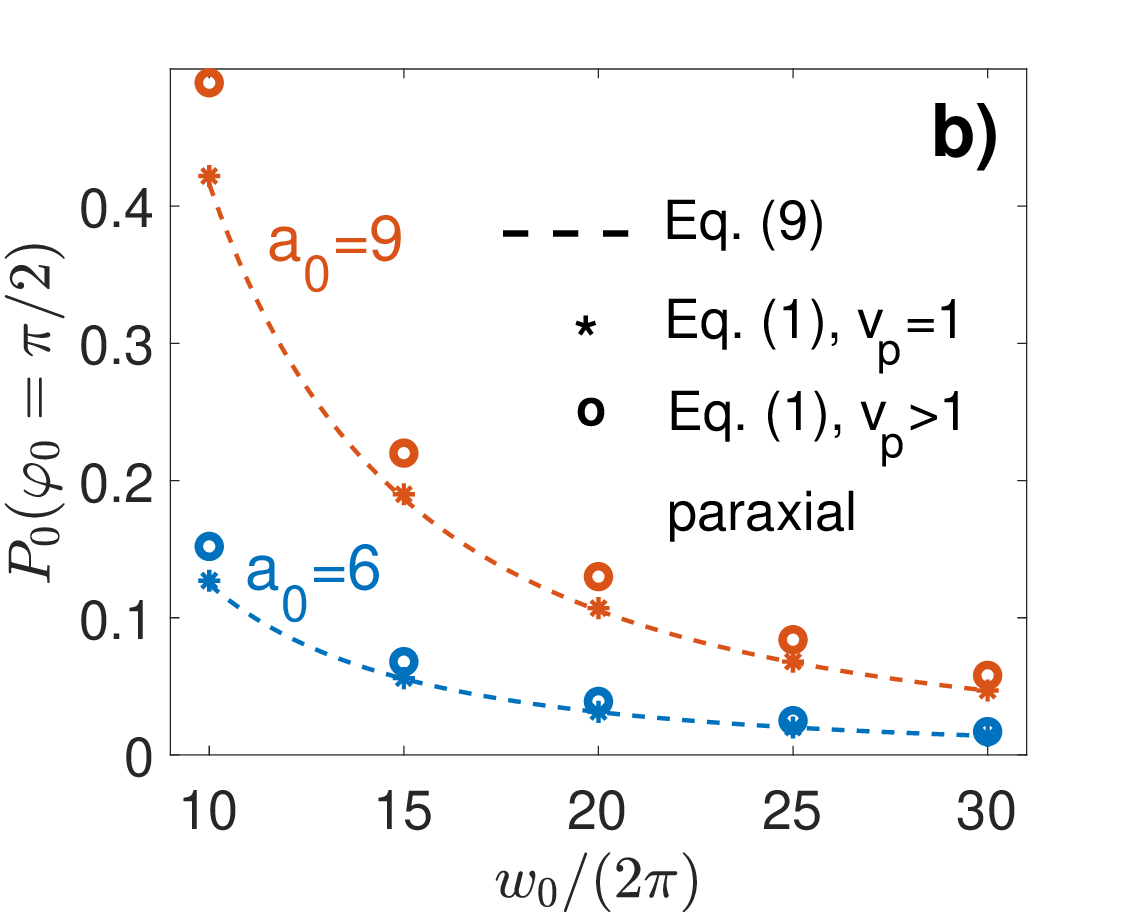}

\includegraphics[width=0.23\textwidth]{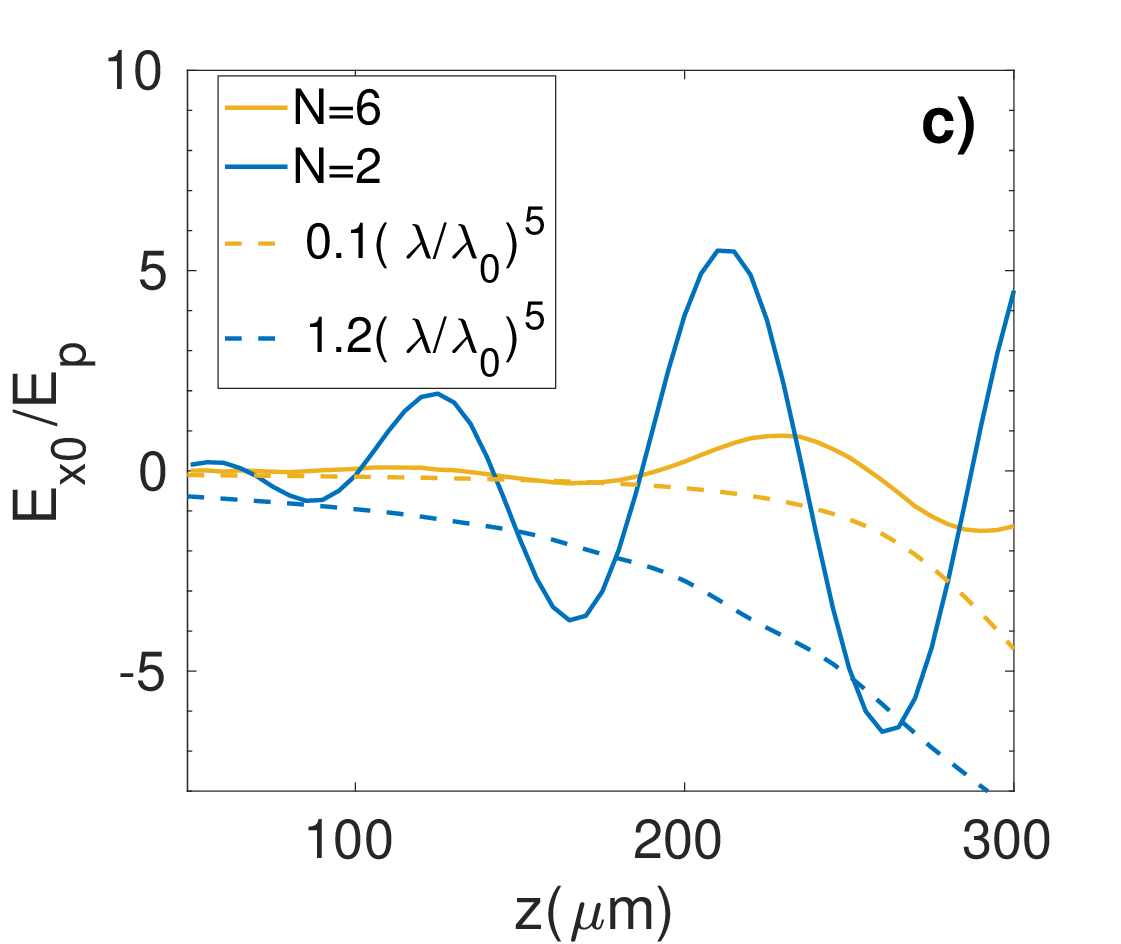}
\includegraphics[width=0.23\textwidth]{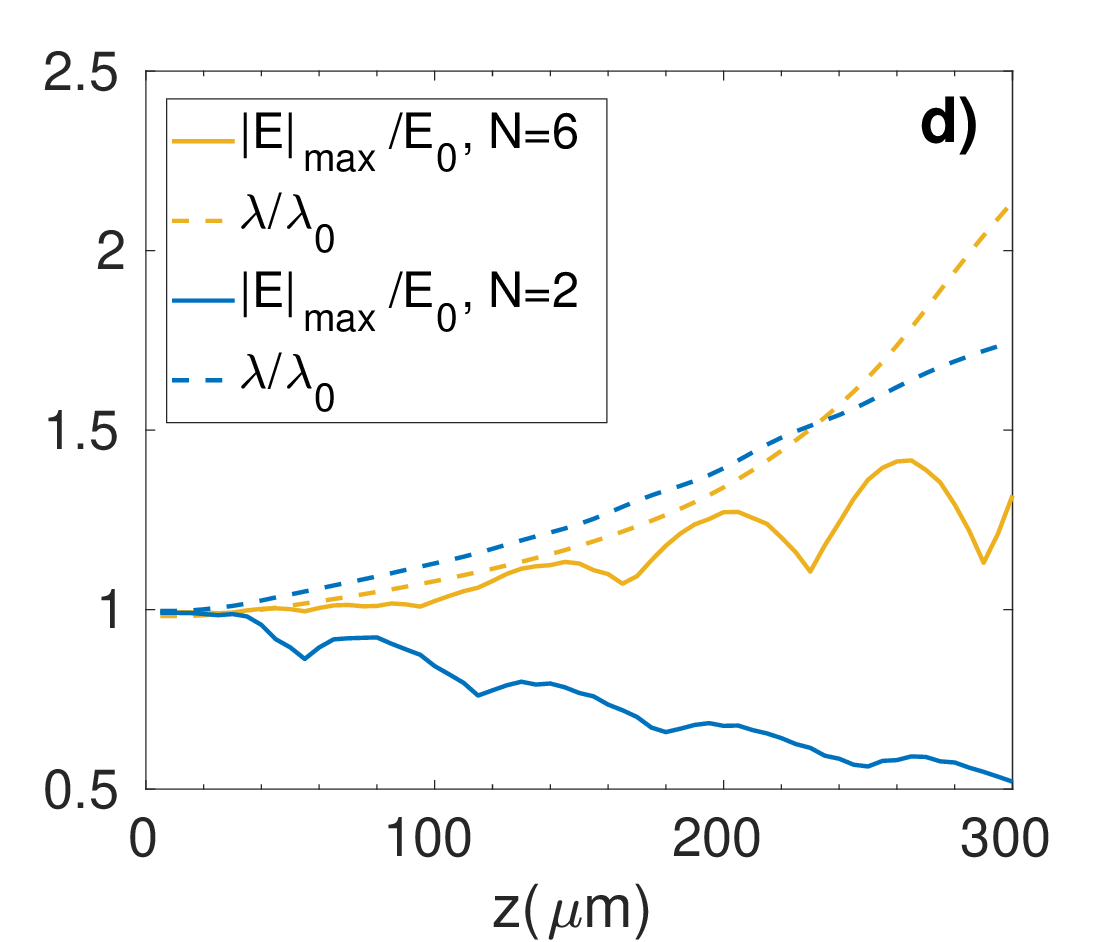}

\caption{ (a) Final momentum of on-axis electrons, given by Eq. (\ref{eq:dpanalit}), after crossing the laser pulse for $a_0=9$ and $w_0=40\pi$. The results of PIC simulations with the same parameters, but varying the CEP phase, are shown with the circles. (b) Comparison of numerical (Eq. (\ref{eq:motion})) and analytical (Eq. (\ref{eq:dpanalit}), dashed lines) results. (c) Transverse electric field measured on axis (full lines) behind the laser pulse in 3D PIC simulation with the parameters presented in the text. The dashed lines show the theoretical prediction, including the variation of the wavelength. (d) Absolute value of the maximum electric field in the laser (full lines) and the variation of the central wavelength (dashed lines) versus propagation distance in the simulations presented in (c). }
\label{Fig3}
\end{figure}

The result of the analytical model (Eq. (\ref{eq:dpanalit})) has been compared to particle-in-cell (PIC) simulations \cite{EXPIC}, where the laser's phase velocity is exact. For valid comparison the laser fields are initialized according to the paraxial description, but non-paraxial corrections can be crucial for realistic few-cycle pulses \cite{nonParax, JOSAB}. The prediction of Eq. (\ref{eq:dpanalit}) is shown in Fig. \ref{Fig3}(a), which underestimates the results of PIC by $\approx$40\%. In Fig. \ref{Fig3}(b) it is proven that Eq. (\ref{eq:dpanalit}) reproduces the numerical result of the equation of motion, given in Eq.(\ref{eq:motion}), for $v_p=1$. However, when a focused laser pulse is considered with $v_p>1$ in the paraxial approximation (circles in Fig. \ref{Fig3}(b)), then $\mathcal{P}_0$ becomes $\approx$25\% larger. In the PIC simulation the phase velocity is not only larger than $c$, but it can vary in space and time \cite{nonParax}, which is probably the reason for the larger discrepancy seen in Fig. \ref{Fig3}(a).

In the case of real laser-plasma interactions the asymmetry increases dramatically because of two reasons: the phase velocity is much larger, compared to the vacuum propagation, and the laser wavelength increases continuously due to longitudinal phase modulation. In this case the asymmetry we characterize with the transverse plasma field on axis ($E_{x0}$) behind the driving pulse, which is proportional to the transverse velocity of electrons, and it should be zero in a symmetric wake. Results of fully self-consistent 3D PIC simulations are presented in Fig. \ref{Fig3}(c) for two pulse lengths ($Sin^2$ envelope) with $w_0=48\pi$ and $a_0=6$, and the plasma density was set to $n_e=10^{19}$ cm$^{-3}$. Here the electric field is normalized to the nonrelativistic wavebreaking field $E_p=m_ec\omega_p/e$, where $\omega_p=(e^2n_e/m_e\epsilon_0)^{1/2}$ is the plasma frequency. At the beginning of the interaction (when $\lambda=\lambda_0 =1 \mu$m) the asymmetry is $10\times$ larger for $N=2$ than for $N=6$, which agrees well with the variation of $C_2$ in Fig. \ref{Fig2}(d). Later, both laser pulses undergo red-shifting, which is shown in Fig. \ref{Fig3}(d). The oscillation period of $E_{x0}$ and of $|E|_{max}$ agrees well with the period of the CEP phase shift estimated as $L_{CEP}\approx n_c/n_e\lambda_0\approx 100 \mu$m, where $n_c=m_e\epsilon_0\omega_0^2/e^2$ is the critical density. 

According to Eq. (\ref{eq:dpanalit2}) the wakefield asymmetry has a very strong dependence on the wavelength: $\mathcal{P}_0\sim (\lambda/\lambda_0)^5$, because the normalized quantities scale as $w_0\sim \lambda_0/\lambda$ and $a_0\sim \lambda/\lambda_0$. We note here that the normalization of the vector potential does not involve the wavelength, but the integral of the laser's electric field, which is the physical observable in the simulation, depends on the wavelength. This power low dependence ($\sim \lambda^5$) is confirmed in Fig. \ref{Fig3}(c), where the dashed lines follow the amplitude of $E_{x0}$.

In conclusion we can say that the exact analytical description of the wakefield asymmetry is only possible when the spectral modulation (red shifting) of the laser pulse is known. When single particle motion is considered the transverse momentum of electrons located on axis can be exactly calculated if the laser's phase velocity in vacuum is known. The analytical formula, Eq. (\ref{eq:dpanalit}), derivd here is valid when $w_0>a_0\lambda_0/(2\pi)$, in standard units. We have pointed out the important role of phase velocity in the evaluation of the ponderomotive force and, more importantly, it has been shown that the maximum value (on axis) of the transverse momentum does not depend on the laser's scalar potential. We also emphasize the crucial importance of the spectral modulation (red-shifting) within the laser pulse, which could help in the correct interpretation of simulations or experiments done with high-power lasers. 

\hspace{10mm}

\begin{acknowledgments}
We acknowledge KIF\"U/NIIF for awarding us access to HPC resource based in Debrecen, Hungary. The ELI-ALPS project (GINOP-2.3.6-15-2015-00001) is supported by the European Union and co-financed by the European Regional Development Fund.
\end{acknowledgments}


\end{document}